\documentclass{emulateapj}
\usepackage{apjfonts}
\usepackage{natbib}
\usepackage{graphicx}
\usepackage{amssymb,amsmath}

\shorttitle{Perturbative Analysis of Synchrotron Spectral Index}

\begin{document}

\title{A Perturbative Analysis of Synchrotron Spectral Index Variation over 
Microwave Sky}

\author{Rajib Saha\altaffilmark{1}, Pavan Kumar Aluri\altaffilmark{2}}

\altaffiltext{1}{Physics Department, Indian Institute of Science 
Education and Research Bhopal,  Bhopal, M.P, 462023, India.} 
\altaffiltext{2}{Inter-University Centre for Astronomy and Astrophysics, 
Post Bag 4, Ganeshkhind,   Pune 411007, India.} 

\begin{abstract}
In this paper, we implement a perturbative approach, first proposed 
by~\cite{Francois1999}, to estimate variation of spectral index of 
galactic polarized synchrotron emission, using linear combination of 
simulated Stokes Q polarization maps of selected frequency bands from 
{\it WMAP} and {\it Planck} observations on a region of sky dominated 
by the synchrotron Stokes Q signal. We find that, a first order 
perturbative analysis recovers input spectral index map well. Along 
with the spectral index variation map our method provides a fixed reference 
index, $\hat \beta_{0s}$, over the sky portion being analyzed. Using Monte 
Carlo simulations we find that, $\langle \hat \beta_{0s}\rangle  = -2.84 \pm 0.01$, 
which matches very closely with position of a peak at $\beta_s(p) = -2.85$, 
of empirical probability density function of input synchrotron indices, 
obtained from the same sky region. For thermal dust, mean recovered spectral 
index, $\langle \hat \beta_d \rangle = 2.00 \pm 0.004$, from simulations,
matches very well with spatially fixed input thermal dust spectral 
index $\beta_d = 2.00$. As accompanying results of the method we also 
reconstruct CMB, thermal dust and a synchrotron template component with 
fixed spectral indices over the {\it entire} sky region. We use full 
pixel-pixel noise covariance  matrices of all frequency bands, estimated 
from the sky region being analyzed, to obtain reference spectral 
indices for synchrotron and thermal dust, spectral index variation map, 
CMB map, thermal dust and synchrotron template components. The perturbative 
technique as implemented in this work has the interesting property that 
it can build a model to describe the data with an arbitrary but enough 
degree of accuracy (and precession) as allowed by the data. We argue 
that, our method of reference spectral index determination, CMB map, 
thermal dust and synchrotron template component reconstruction is a maximum 
likelihood method.   
\end{abstract}

\keywords{cosmic background radiation --- cosmology: observations --- diffuse radiation}
\maketitle

\section{Introduction}

Observations of microwave signal by {\em WMAP} and {\em Planck}
satellite missions, in the frequency range $23$ GHz to $857$ GHz, 
can be used to obtain a wealth of precise cosmological information 
(e.g, precise values of basic cosmological parameters~\citep{WMAPCosmoParam2013,
WMAPCosmoResults2013, PlanckCosmoParam2015}, topology 
of space~\citep{PlanckTopology2015}, laws and nature of 
physical conditions existing at different epoch marked by 
different phases of dynamical evolution of space-time
~\citep{PlanckInflation2015, PlanckNG2015, PlanckIsotropy2014, 
PlanckPrimMagField2015}) as well as precise information 
about various astrophysical mechanisms operating since recent 
times that cause diffuse emissions inside Milky Way~\citep{Bennett03_fg, 
PlanckLFIFg2015,PlanckHaze2013, HazeBubble2012}. However, 
extracting these information accurately with precision requires a 
 method that separates different components from their observed 
mixture using a model that is flexible enough to describe the data 
with an arbitrary but desired degree of minute closeness {\it as permissible by data}. 
A completely accurate model may produce ambiguous results if the
data is simply insufficient  to constrain all model parameters correctly 
e.g., consider the astrophysical problem of reconstruction 
various galactic emission components which is a greatly complex task
due to inherent degeneracy existing in the morphological pattern 
of these emissions, degeneracy arising due to similarities in 
physical models of multiple emissions or due to other reasons.   
Such uncertainties in reconstructed foreground components may
cause cosmological inferences drawn from the analysis of 
reconstructed Cosmic Microwave Background (CMB) maps to be biased~\citep{Tegmark2000} 
if these uncertainties  are not known or if they are not properly 
taken into account. 
 
One of the prime source of uncertainties against properly extracting 
the primordial signal and other galactic foreground components from 
observed maps is {\em intrinsic variation of spectral index 
of synchrotron emission}~\citep{Tegmark1998} over the sky due to
variations of number density of cosmic ray electrons in our galaxy.  
In this paper we implement a novel method on simulated {\em WMAP} 
and {\em Planck} Stokes Q frequency maps to perturbatively model 
the synchrotron spectral index variation over the sky following 
the approach proposed by~\cite{Francois1999}. The perturbative 
nature of the method is very desirable since it allows one to build 
a model with just enough accuracy that is still meaningful within the 
available data. As associated results
of the method we reconstruct best-fit  map for CMB and templates for
thermal dust and synchrotron emission.

Usefulness of the method described in this paper can also 
be seen from a different perspective.  There exist several methods 
in the literature for CMB or (and) foreground component reconstruction  
which can be grouped broadly in following three categories:

\begin{enumerate}
\item{The so-called {\it blind, internal-linear-combination} 
approach,~(\citep{Tegmark96, Maino2002, 
Tegmark2003, Delabrouille2009, Saha2011,Bennett2012,Planck2015_CMB}),
where one removes foregrounds based upon the fact that 
CMB follows blackbody spectrum whereas, other components 
do not.  In principle, this method can also be used to produce a  
cleaned map of any of the foreground components~(\cite{Remazeilles2011}), 
provided its spectral index is {\it a priori} known and is constant 
over  the sky portion being analyzed. The method by definition 
is a non-Maximal Likelihood (ML) approach since it incorporates 
minimum information about the physical components and no information about 
detector noise.}

\item{The second class~(\cite{Eriksen2008a,Eriksen2008b,Planck2015_fg,Planck2015_CMB}) 
is that one where one uses as much prior information as possible 
about all components, along with the detector noise model, to 
perform a ML analysis of the data. The outcome of 
this method are ML estimates of all components. 
}

\item{The third one is an intermediate approach of the first and second 
categories. Although, here one incorporates informations about 
various components  available from prior observations, (e.g., 
Maximum Entropy Method,~\citep{Hobson1998, Bennett03_fg,Hinshaw_07,
Gold2009,Gold2011}~\footnote{See also all but first one of these 
papers for analysis following methods belonging to first and second 
categories.} and the Wiener filtering approach,~\citep{Bunn1994, 
Tegmark96, Francois1999a}),  it is a non ML approach. 
} 
\end{enumerate}

All the above methods have merit and demerit of 
their own. ML approach has advantages that it 
is the {\it optimal} method in the sense that it retains 
all information of the data and provides maximum precision 
on the derived parameters. ML method may, however, be biased 
if the model to be used for the data is not accurately known, 
or, if there are degeneracies between the parameters.  The 
non-ML approaches are advantageous since data is processed 
with the minimal assumptions. However, it is a {\it sub-optimal}
method and can be biased. One, therefore, 
desires to ask a question {\it can we develop a method which 
requires minimal assumptions in terms of data model (e.g., no 
requirement of using templates) and, at the same time, 
which is a ML approach with respect to at the least almost 
all of its parameters?} So far, there has not been any attempt 
in the literature to answer this question. The novel perturbative approach 
for component separation and for determination of  variation of 
spectral index of synchrotron emission followed in this  paper, 
at one-side, is a  linear-combination method, on the other side, 
is a ML approach with respect to all of its component templates
and best-fit spectral indices for synchrotron and thermal dust. 
The method incorporates  information about  spectral properties 
of components and detector noise model, thus preserving properties 
of first and second categories  described above.   

There have many attempts in the literature to measure synchrotron 
spectral index of our Galaxy. \cite{Reich1988} find map of 
synchrotron spectral index between $420$ MHz and $1420$ MHz using radio 
continuum surveys. Following the method of TT plot~\cite{Davies1996} 
find synchrotron temperature spectral index lies in the range $-2.8$ 
to $-3.2$ from a strip at $\delta=40^{\circ}$ latitude sections
of $408$ MHz and $1420$ MHz maps.~\cite{platania97} find synchrotron 
spectral index between $420$ MHz and $7.5$ GHz as $-2.76 \pm 0.11$.
\cite{Bennett03_fg} find a flatter synchrotron spectral index, 
$\sim -2.5$, near the galactic plane and a steeper index, $\sim -2.9$ 
towards the higher galactic latitude.  Using {\it WMAP} nine-year data
~\cite{Fuskeland2014} estimate polarized synchrotron spectral 
index towards galactic plane as $-2.98\pm 0.01$, and a steeper 
spectral index, $-3.12\pm 0.04$, towards the higher  galactic latitude.

We present our paper as follows. We describe the problem in 
Section~\ref{Statement}. In Section~\ref{Solution} we describe 
the perturbative approach used to model the data. We discuss 
basic formalism of this work in Section~\ref{Formalism}. 
Here we discuss 1) the power minimization algorithm (based 
upon the model described in Section~\ref{Solution}) and 
how the power minimization algorithm leads to an equation 
for weights that we use for linear combination of input Stokes 
Q maps, later in Section~\ref{Methodology}, 2) the difference map 
at each frequency bands, and the noise covariance matrix thereof, 
3) the $\chi^2$ data-model mismatch statistics and 4) relationship 
of our method with ML approaches. In Section~\ref{Methodology} we 
describe the method used in this  paper. We present results in 
Section~\ref{Results}. Finally, we conclude in Section~\ref{Conclusion}.
    
\section{Basic Problem}
\label{Statement}
Let us assume we have observations of CMB and foreground polarizations at 
$n_b$ different frequency bands. Each frequency band 
contains contribution from $n$ different physical components, namely, CMB  and 
emissions from $n-1$ foreground components - each with a given spectral index
all over the sky - plus detector noise~\footnote{Detector noise 
does not count as a physical component.}.
If foreground component $j$ has a fixed spectral index, $\beta_j$, all over the sky, net 
polarized Stokes Q signal, $Q_i(p)$, at a pixel $p$, due to CMB, $Q_c(p)$, all foregrounds, $Q_{0j}$,
and detector noise, $Q^n_i(p)$ at a frequency $\nu_i$ in thermodynamic temperature 
unit is given by,
\begin{eqnarray}
 Q_i(p) = Q_c(p) + \sum_{j=1}^{n-1} a_i
\left( \frac{\nu_i}{\nu_{0j}}\right)^{\beta_j} Q_{0j}(p) + 
 Q^n_i(p) \, ,
\label{WE0} 
\end{eqnarray}
where  $a_i \equiv a(\nu_i)$ denotes conversion factor from antenna to 
thermodynamic temperature unit at frequency $\nu_i$ for all foreground 
components. CMB anisotropy at any  given pixel $p$ is independent 
on frequency $\nu_i$ due to its blackbody nature. Each foreground 
template $Q_{0j}$ is defined with reference to frequency $\nu_{0j}$,
which may be different for different foreground components. 
We note that, in general, observed maps for different frequencies 
are convolved by different instrument response functions. However, 
we implicitly assume that frequency maps, $Q_i(p)$, in Eqn~\ref{WE0} already 
have been brought to a common resolution by properly deconvolving 
first by respective instrument response functions and then convolving 
again by the common response function. Since all frequency maps of 
our work are convolved by a common Gaussian beam function of FWHM = $20^\circ$
we omit reference to beam function in all equations of this paper containing frequency maps,
CMB or foreground templates. 
 
For polarized signal there are only two foreground components at microwave 
frequencies, namely,  synchrotron and thermal dust. Moreover, the spectral index 
for synchrotron emission varies with sky position due to variation of relativistic 
electron density over the sky.  In the presence of variation of spectral index 
for synchrotron, but fixed spectral index for thermal dust component Eqn~\ref{WE0} 
reduces to,
\begin{eqnarray}
\begin{aligned}
 Q_i(p) = {} &  Q_c(p) + a_i  \left( \frac{\nu_i}{\nu_{0s}}\right)^{\beta_s(p)} Q_{0s}(p) \\
  & + a_i \left( \frac{\nu_i}{\nu_{0d}}\right)^{\beta_d} Q_{0d}(p)  +  Q^n_i(p) \, ,
\label{WE1}
\end{aligned}
\end{eqnarray}
where $\beta_s(p)$ represents the synchrotron spectral index at a pixel $p$ 
and $\beta_d$ is the global thermal dust spectral index. {\it The motivation for our work is to 
obtain a reliable estimate of $\beta_s(p)$ using a perturbative analysis, without 
assuming any prior template models for foregrounds or CMB.}

\section{Perturbative Data Model}
\label{Solution}
Let us consider a given region of sky. Synchrotron spectral index at a 
pixel $p$ can be written as, $\beta_s(p) = \beta_{0s} + \Delta \beta_s(p)$, 
where $\beta_{0s}$ represents some reference spectral index for the region and 
$\Delta \beta_s(p)$ represents actual variation of spectral index over 
$\beta_{0s}$ inside the region. Using the direction dependent spectral index,
$\beta_s(p)$, in the second term of Eqn~\ref{WE1}, one can show that~\citep{Francois1999},
\begin{equation}
\begin{aligned}
{} & \left( \frac{\nu_i}{\nu_{0s}}\right)^{\beta_s(p)} =   
\left( \frac{\nu_i}{\nu_{0s}}\right)^{\beta_{0s}\left(1 + \frac{\Delta \beta_s(p)}{\beta_{0s}}\right)}\\
& \sim \left( \frac{\nu_i}{\nu_{0s}}\right)^{\beta_{0s}} 
\left( 1 + \Delta \beta_s(p)\ln\left(\frac{\nu_i}{\nu_{0s}} \right) \right)\, , 
\label{WE2}
\end{aligned}
\end{equation}
where the second line follows from Taylor expansion of the left 
hand side of above equation up to first order in $\Delta \beta_s(p)/\beta_{0s}$, 
assuming $\Delta \beta_s(p)/\beta_{0s} \ll 1$. 
Using Eqn.~\ref{WE2} in Eqn.~\ref{WE1} we  obtain, 
\begin{eqnarray}
\begin{aligned}
 Q_i(p){} & {}  =  Q_c(p) +  a_i\left(\frac{\nu_i}{\nu_{0s}}\right)^{\beta_{0s}} Q_{0s}(p) 
   + a_i\left(\frac{\nu_i}{\nu_{0s}}\right)^{\beta_{0s}} \ln\left(\frac{\nu_i}{\nu_{0s}} \right)\\
& \times  Q_{0s}(p)\Delta \beta_s(p) + 
a_i \left( \frac{\nu_i}{\nu_{0d}}\right)^{\beta_d} Q_{0d}(p)  + Q^n_i(p) \, .
\label{WE3}
\end{aligned}
\end{eqnarray}
Apart from CMB and detector noise, our data model given by above equation, 
can now be interpreted nicely to consist of a total of three foreground 
emission components, namely, synchrotron with a constant spectral index 
$\beta_{0s}$, a second component $Q_{0s}(p) \Delta \beta_{s}(p)$, 
with its frequency variation 
$a_i\left( \frac{\nu_i}{\nu_{0s}}\right)^{\beta_{0s}}\ln\left(\frac{\nu_i}{\nu_{0s}} \right)$
and finally thermal dust component. As we can see from above equation our 
model renders the variation of spectral index of synchrotron component in 
terms of a new foreground component with rigid frequency scaling. We define 
a set of templates, $T_j$, where $j \in \{1,2,3,4\}$ following,
\begin{eqnarray}
\begin{aligned}
& T_1(p) =  Q_c(p)\\
& T_2(p) =  Q_{0s}(p)\\
& T_3(p) =  Q_{0s}(p)\Delta \beta_s(p)\\
& T_4(p) =  Q_d(p) \, .
\end{aligned}
\label{templates}
\end{eqnarray}
With these definitions, total polarized emission of Eqn.~\ref{WE3},
at a frequency $\nu_i$ is then given by,
\begin{eqnarray}
 Q_i(p) = \sum_{j=1}^{4}T_j(p)s^j_i +  Q^n_i(p) \, ,
\label{Comp1}
\end{eqnarray}
where we have denoted emission from $j^{th}$ component at channel $i$ 
(in thermodynamic temperature unit) by $T_j(p)s^j_i$. $T_j(p)$ denotes  
template for $j^{th}$ component based on reference frequency $\nu_{0j}$ 
and $s^j_i$, for a given $j$, denotes elements of so-called shape vector, 
$[{\bf s}]_j$, for component $j$. For foregrounds with emissions following 
usual power law behavior elements of the shape vector are given 
by~\footnote{For CMB elements of shape vector are unity for 
all frequency bands and choice of $\nu_{0c}$ is irrelevant.},
\begin{eqnarray}
s^j_i = \left(\frac{\nu_i}{\nu_{0j}}\right)^{\beta_j}a(\nu_i) \, ,
\end{eqnarray}
where $\beta_j$ represents spatially fixed spectral index of $j^{th}$ 
(foreground) component. Following method as described in Section~\ref{Methodology} 
we reconstruct each template defined by Eqn~\ref{templates}. Due to presence 
of detector noise, the reconstructed templates, $\hat T_j(p)$, are, however, different 
from $T_j(p)$. Once all template components are reconstructed, synchrotron 
spectral index variation map is obtained following, 
\begin{equation}
\Delta \hat \beta_s(p) = \hat T_3(p)/\hat T_4(p) \, .
\label{SpIndex}
\end{equation}

\section{Formalism}
\label{Formalism}
\subsection{Component Reconstruction and Weights}
From a mixture of $n$ physical components present in different frequency bands, our 
aim is to isolate  best-fit map of each one of them by removing rest. This can be achieved by 
reconstructing any one component first and then repeating the method for all other components. 
Let us denote the component to be reconstructed by a Greek letter subscript, e.g., $\alpha$  and other 
components to be removed by any Roman letter subscript taking values from the difference set 
$\{1,2,...,n\}-\{\alpha\}$. 
We denote shape vector for the component to be 
reconstructed by $[{\bf s}]_{\alpha}$ while each of the rest $n-1$ components has shape 
vector $[{\bf s}]_i$. 
To reconstruct the cleaned template, $\hat T_{\alpha}(p)$, for component $\alpha$, we  
form a linear combination of available maps following,
\begin{eqnarray}
\hat T_{\alpha}(p) =  \sum_{i=1}^{n_b}w^i_{\alpha} Q_i(p) \, ,
\end{eqnarray}
where the linear weights, $w^i_{\alpha}$, for $\alpha$ component is obtained by minimizing the 
total pixel-pixel variance, $\sigma^2_{\alpha}$, of the cleaned template, $\hat T_{\alpha}(p)$. 
Pixel variance of of this template is defined as,
\begin{eqnarray}
\sigma^2_{\alpha} = \sum_{p=1}^{\mathcal N_{\textrm{pix}}}\left( \hat T_{\alpha}(p)
 - \sum_{p'}\frac{\hat T_{\alpha}(p')}{\mathcal N_{\textrm{pix}}}) \right)^2 \, ,
\end{eqnarray}
where $\mathcal N_{\textrm{pix}}$ denotes total number of  pixels available for analysis in a single frequency map after
any suitable mask has been applied . Using matrix notation, above equation 
can easily be written as,
\begin{eqnarray}
\sigma^2_{\alpha} = [{\bf W}]_{\alpha}[{\bf V}][{\bf W}]^T_{\alpha} \, , 
\end{eqnarray}
where, $[{\bf W}]_{\alpha}$ is a $1\times n_b$ weight matrix, 
$(w^1_{\alpha}, w^2_{\alpha},....,w^{n_b}_{\alpha})$, for component $\alpha$. 
$[{\bf V}]$ represents, $n_b\times n_b$  covariance matrix, 
with elements $V_{ii'} $ representing covariance between frequency maps $Q_i(p)$ and 
$Q_{i'}(p')$. As described in details in~\cite{Saha2008} the choice of 
weights which minimizes $\sigma^2_{\alpha}$ is given by, 
\begin{eqnarray}
[{\bf W}]_{\alpha} = \frac{[{\bf s}]_{\alpha}[{\bf V}]^+}{[{\bf s}]_{\alpha}[{\bf V}]^+[{\bf s}]^T_{\alpha}} \, ,
\label{weights}
\end{eqnarray}
where $[{\bf V}]^+$ represents Moore-Penrose generalized inverse~\citep{Moore1920,Penrose1955} of $[{\bf V}]$.  
Our aim is to obtain an analytical expression for weights that minimizes variances 
due to physical components in the limiting situation when detector noise is negligible~\footnote{The 
assumption of limiting detector noise is justified for our current work, since we use low 
resolution maps for our analysis.}.
Using variance and covariance of different components we can write, 
\begin{eqnarray}
[{\bf V}] = \sigma^2_{\alpha}[{\bf s}]^T_{\alpha}[{\bf s}]_{\alpha} + [{\bf r}]^T_{\alpha}[{\bf s}]_{\alpha} 
+ [{\bf s}]^T_{\alpha}[{\bf r}]_{\alpha} +  [{\bf F}] \, ,
\end{eqnarray}
where $[{\bf r}]_{\alpha}$ is a $1\times n_b$ matrix describing random chance correlation 
between the component to be reconstructed and all of the other components at each of the 
$n_b$ frequency bands. $ [{\bf F}]$ describes the $n_b \times n_b$ covariance matrix 
of $n-1$ components, hence it has a rank $n-1$. Above equation can be written as,
\begin{eqnarray}
[{\bf V}] = \sigma^2_{\alpha}[{\bf s}]^T_{\alpha}[{\bf s}]_{\alpha} + [{\bf A}]_1 \, ,
\end{eqnarray}
where $ [{\bf A}]_1 = [{\bf r}]^T_{\alpha}[{\bf s}]_{\alpha}  + [{\bf A}]_2$ and $[{\bf A}]_2 = 
 [{\bf s}]^T_{\alpha}[{\bf r}]_{\alpha} +  [{\bf F}]$. Using above equation we can apply generalized 
Shermann-Morisson formula~(\cite{JKB,CDM}) successively to obtain both numerator 
and denominator of Eqn~\ref{weights} (e.g., as in~\cite{Saha2008}). After a long algebra 
and neglecting all terms containing $[{\bf r}]_{\alpha}$, Eqn~\ref{weights} reduce to 
\begin{eqnarray}
[{\bf W}]_{\alpha} = \frac{[{\bf s}]_{\alpha}([{\bf I}]-[{\bf F}][{\bf F}]^+)}{[{\bf s}]_{\alpha}([{\bf I}]-
[{\bf F}][{\bf F}]^+)[{\bf s}]^T_{\alpha}} \, .
\label{weights2}
\end{eqnarray}
The term $[{\bf F}][{\bf F}]^+$ can be interpreted as the projector on the column 
space of $[{\bf F}]$. Hence $([{\bf I}]-[{\bf F}][{\bf F}]^+)$ is the orthogonal 
projector on the null space of  $[{\bf F}]$. We note that, $[{\bf s}]_i \in 
\mathcal C([{\bf F}])$, for $i=\{1,2,3,...,n\}-\{\alpha\}$. Hence, $[{\bf F}][{\bf F}]^+ 
[{\bf s}]^T_i= [{\bf s}]^T_i$ i.e., $[{\bf W}]_{\alpha}[{\bf s}]^T_i =0$,  however, since 
$[{\bf s}]_0 \notin \mathcal C([{\bf F}])$ we have $[{\bf W}]_{\alpha}[{\bf s}]^T_{\alpha} =1$, 
which is the condition for reconstruction of the component under consideration. 
Eqn.~\ref{weights2} is not directly usable by us since {\it a priori} we do not have 
any knowledge about the templates to be reconstructed from data and hence about the 
covariance matrix $[{\bf F}]$. We, however, see that the crucial properties 
of $[{\bf W}]_{\alpha}$, viz., $[{\bf W}]_{\alpha}[{\bf s}]^T_i =0$ and 
$[{\bf W}]_{\alpha}[{\bf s}]^T_{\alpha} =1$ that must be satisfied for reconstruction of
$\alpha$ component remain unchanged if we replace $[{\bf F}]$ in Eqn~\ref{weights2} by a 
new $n_b \times n-1$ matrix, $[{\bf G}]$,  whose column vectors are given by each of 
the shape vectors of the $n-1$ components which are to be removed from the data. We therefore 
compute weights replacing $[{\bf F}]$ by $[{\bf G}]$ in Eqn~\ref{weights2}. 

\subsection{Difference Map and Noise Covariance}
Once all components have been reconstructed, using weights following 
Eqn~\ref{weights2}, for the complete set of possible shape vectors 
of all components, we can form reconstructed frequency map, $\hat Q_i(p)$,  at each 
frequency, $\nu_i$, following,
 \begin{eqnarray}
\hat Q_i(p) = \sum_{j=1}^{n}\hat T_j(p)s^j_i \,.
\label{FreqMapRec}
\end{eqnarray}
Due to underlying linearity in our method, one can clearly see 
that  noise, $\tilde Q^n_k(p,\nu_i)$, in the  $k^{th}$ reconstructed template 
at frequency $\nu_i$,  has contribution from noise of all input frequency maps. Specifically,
\begin{eqnarray}
\tilde Q^n_k(p,\nu_i) = s^k_i\sum^{n_b}_{j=1}w^j_k Q^n_j(p)\, .
\end{eqnarray}
Hence the total noise, $\tilde Q^n_i(p)$, due to all recovered components  at frequency $\nu_i$, is given by,
\begin{eqnarray}
\begin{aligned}
\tilde Q^n_i(p) = {} & \sum^{n}_{k=1} \tilde Q^n_k(p,\nu_i) = 
\sum^{n}_{k=1}s^k_i\sum^{n_b}_{j=1}w^j_k Q^n_j(p)\\
 & =\sum^{n_b}_{j=1}\tilde w^j_i Q^n_j(p) \, , 
\label{Noise1}
\end{aligned}
\end{eqnarray}
where we have defined, 
\begin{eqnarray}
\tilde w^j_i = \sum^n_{k=1}s^k_iw^j_k \,.
\end{eqnarray}
We form a difference map at each frequency, $\nu_i$, by subtracting the 
recovered and input frequency map at the same frequency, $\nu_i$. Using 
Eqn~\ref{Noise1} noise in the $i^{th}$ difference map is given by,
\begin{eqnarray}
Q^n_i(p)- \tilde Q^n_i(p) = (1-\tilde w^i_i)Q^n_i(p) -  
\sum^{n_b}_{j\ne i}\tilde w^j_i Q^n_j(p)\, . 
\end{eqnarray}
Assuming  noise of different frequency bands have zero mean and noise properties 
of different detectors are uncorrelated with one another, noise covariance of the 
$i^{th}$ difference map is given by,
\begin{eqnarray}
\begin{aligned}
N^{i,D}_{pp'}= {} & (1-\tilde w^i_i)^2\left<Q^n_i(p)Q^n_i(p')\right> + 
\sum^{n_b}_{j\ne i}\left(\tilde w^j_i\right)^2 \left< Q^n_j(p)  Q^n_j(p')\right> \\
 & =(1-\tilde w^i_i)^2 N^i_{pp'}+ \sum^{n_b}_{j\ne i}\left(\tilde w^j_i\right)^2N^j_{pp'} \, ,
\label{CovDiffMap}
\end{aligned}
\end{eqnarray} 
where $N^{i(j)}_{pp'}$ is the noise covariance matrix of the $i^{th}(j^{th})$ input 
frequency map.

\subsection{Data-Model Mismatch Statistic}
Using the noise covariance matrix obtained in the previous section we define  mismatch 
between the  data and model by means of usual $\chi^2$ statistic following,
\begin{eqnarray}
\begin{aligned}
\chi^2  = \frac{1}{N}\sum_{i,p,p'}\left( Q_i(p)-
 \tilde Q_i(p)\right)\left( N^{i,D}\right)^{+}_{pp'} 
 \left(Q_i(p')-\tilde Q_i(p')\right) \, ,
\label{chi2}
\end{aligned}
\end{eqnarray}
where $(N^{i,D})^{+}_{pp'}$ represents 
Moore-Penrose generalized inverse of $N^{i,D}_{pp'}$ and $N = n_b\times \mathcal N_{\textrm{pix}}$ denotes total 
number of surviving pixels counting all frequency bands after any suitable mask has 
been applied on the maps. Since, $\tilde Q_i(p)$ depends upon $\beta_{0s}$, $\beta_d$ 
through Eqn~\ref{FreqMapRec} and $N^{i,D}_{pp'}$ depends upon $\beta_{0s}$, $\beta_d$ and 
frequency band noise covariance matrices through Eqn~\ref{CovDiffMap}, we have 
$\chi^2 \equiv \chi^2(\beta_{0s},\beta_d)$. Clearly,  $\beta_{0s}$ and $\beta_d$ enter as the minimizing 
variables for the $\chi^2$ statistic. Our best fit values for these spectral indices are 
the ones which together minimizes the $\chi^2$ given by above equation. 

\subsection{Relationship with ML approaches}

Eqn~\ref{chi2} shows that the best fit spectral indices are ML estimates for a known detector noise 
model provided we have also chosen uniform prior on the indices. The ML nature of these indices directly
implies that all the reconstructed templates, $\hat T_j(p)$, are also ML estimates. We, however, note that our
reconstructed spectral index map is not a ML estimate since we construct it following Eqn~\ref{SpIndex},
by taking ratio of two templates, without directly estimating it from the fit.

\begin{deluxetable}{ccccc}
\tabletypesize{\small}
\tablewidth{0pt} 
\tablecaption{Frequency maps }
\tablehead{\colhead{Frequency, $\nu$}          &
           \colhead{{\em WMAP}, {\em Planck} }             &
           \colhead{Name}                      &
           \colhead{$a(\nu)$}                  &
           \colhead{$\sigma_0 (\mu K)$}         \\
           \colhead{(GHz)}              &        
           \colhead{}                          &
           \colhead{}                          &
           \colhead{}                          &               
           \colhead{Ref.\,\tablenotemark{a}}   }
\startdata
$23$   & {\em WMAP}            &  K      & $1.014$     & $1435.0$ \\
$30$   & {\em Planck}, LFI     &  A      & $1.023$     & $36.56$  \\  
$33$   & {\em WMAP}            &  Ka     & $1.029$     & $1472.0$ \\
$41$   & {\em WMAP}            &  Q(1,2) & $1.044$     & $2254.0$, $2140.0$\\
$44$   & {\em Planck}, LFI     &  B      & $1.051$     &$37.03$ \\
$61$   & {\em WMAP}            &  V(1,2) & $1.099$     &$3324.0$,  $2958.0$ \\ 
$71$   & {\em Planck}, LFI     &  C      & $1.136$     &$37.11$\\
$100$  & {\em Planck}, HFI     &  D      & $1.283$     &$15.83$\\
$143$  & {\em Planck}, HFI     &  E      & $1.646$     &$11.80$\\
$217$  & {\em Planck}, HFI     &  F      & $2.961$     &$19.39$
\enddata
\tablenotetext{a}{Stokes (Q, U) polarization noise level in 
$\mu K$ thermodynamic temperature unit at  $N_{\textrm{side}} = 512$.}
\tablecomments{{\ \ } List of {\emph WMAP} and {\emph Planck} 
frequency maps used in the analysis. Wherever multiple differencing 
assemblies are available for {\emph WMAP} polarized noise level 
for each differencing assemblies is also mentioned in order. 
{\em Planck} noise levels are consistent with the {\em Planck} 
blue book~\citep{BlueBook2005} values and scaled to $N_{\textrm{side}} = 512$. For {\em WMAP} 
actual noise level per pixel at $N_{\textrm{side}} = 512$ is given 
by $\sigma(p) = \sigma_0/\sqrt{N_{\textrm{obs}}(p)}$, where $N_{\textrm{obs}}(p)$ 
denotes effective number of observations at a pixel $p$ for 
given DA map. For {\em Planck} bands noise level per pixel 
at $N_{\textrm{side}} = 512$ is assumed to be uniform with  $\sigma(p)=\sigma_0$ 
as the specified values at $N_{\textrm{side}}=512$.}  
\label{ListMaps}
\end{deluxetable}

\section{Methodology}
\label{Methodology}
\subsection{Input Data}
\subsubsection{Frequency bands}
We include simulated Stokes Q polarization maps of {\em WMAP} K, Ka, Q, V bands, all 
three {\em Planck} LFI bands, and three lowest {\it Planck} HFI bands. We 
do not include {\em WMAP} W band in our analysis since it contains somewhat 
larger noise level compared to other {\it WMAP} bands. We do not include 
two highest frequency {\em Planck} HFI maps  since they are strongly dominated by thermal 
dust component. Moreover, with inclusion of HFI $353$ GHz  map in our analysis, 
$\chi^2$ values become increasingly more sensitive to  thermal dust model, 
giving lesser weightage to synchrotron component. Hence we exclude this 
frequency band also from our analysis. We however verify  that conclusions drawn in this 
paper from spectral index analysis remain unchanged with inclusion of HFI 
$353$ GHz map.  A detailed specifications of the frequency bands used 
in this work is given in Table~\ref{ListMaps}.

Since pixel values of spectral index variation map, $\Delta \beta_s(p)$, are 
small compared to the foreground emission, any significant residual noise in the 
reconstructed template maps results in its potentially noisy reconstruction 
following Eqn~\ref{SpIndex}. Owing to delicate noise sensitivity of reconstructed 
spectral index map, we chose to minimize detector noise at the beginning of analysis,
by first downgrading all Stokes Q frequency maps at $N_{\textrm{side}} = 8$ and 
then smoothing each one by a Gaussian beam function of FWHM = $20^\circ$.

\begin{figure}[t]
\includegraphics{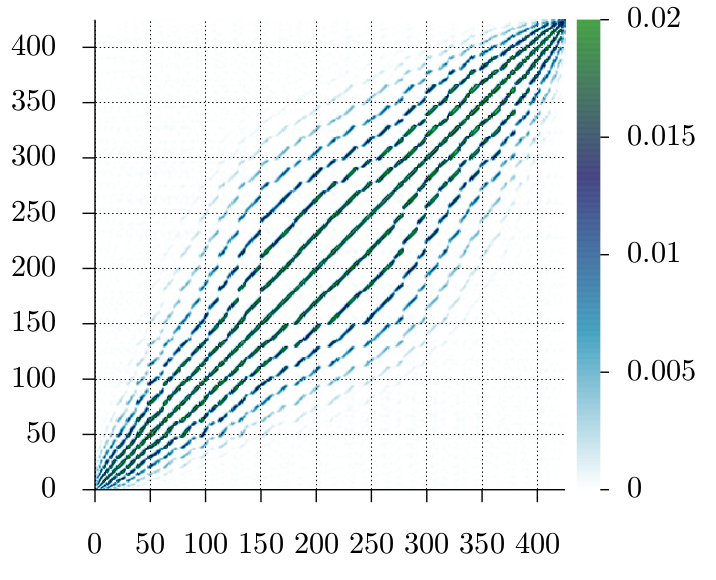}
\caption{Full pixel-pixel noise covariance matrix obtained from $N_{\textrm{nside}} = 8$,
$20^{\circ}$ smoothed maps over surviving pixels of {\it pm10} mask for V band. }
\label{CovMat}
\end{figure}

\subsubsection{Mask}
We see from Eqn~\ref{SpIndex}, $\Delta \hat \beta_s(p)$ map may contain 
noisy pixels  if corresponding pixels in the reconstructed synchrotron 
template, $\hat T_4(p)$, (or $\hat T_3(p)$), are also noise dominated. 
To avoid noisy reconstruction of spectral index map, we remove those pixels 
from all $10$ input frequency maps which contain weak synchrotron Stokes Q polarization 
signal. For this purpose, based upon the smoothed Stokes Q  synchrotron 
template at K band, we make a mask which excise  all pixels with absolute 
values less than 10$\mu K$ in antenna temperature unit. We call this mask {\it pm10} mask 
and apply this mask to all frequency maps for reconstruction of spectral index
map~\footnote{We apply {\it pm10} mask only for 
reconstruction of spectral index map. We present results for reconstruction of CMB and foreground  components 
(e.g., Section~\ref{CMBForeground}) over entire sky.}. 

\subsubsection{Physical Components}
\label{ICompSpIndex}
We use {\em WMAP} `base fit' MCMC (Markhov Chain Monte Carlo)  
synchrotron and thermal dust polarization maps for our simulations. 
These maps are provided  at $N_{\textrm{side}} = 64$ at LAMBDA 
website. As mentioned at LAMBDA website, for some of the pixels 
of these maps, MCMC fitting resulted in errors. For both Q and 
U polarized synchrotron map at K band, using {\em WMAP}'s bit-coded 
error flag file, we first identify all pixels for which {\em WMAP} 
MCMC fit resulted in error. The process identifies a total of $1315$ pixels 
at $N_{\textrm{side}} = 64$. Pixel value for each of these bad pixels is then replaced 
by the average pixel value of its  nearest neighbors. If a bad pixel has 
one or more bad neighbours we average only over the good ones.  
Following similar procedure we obtain polarized thermal dust template at W band at 
$N_{\textrm{side}} = 64$. We downgrade each of these two templates to $N_{\textrm{side}} = 8$. 
We generate polarized CMB maps compatible to WMAP LCDM spectrum directly at 
$N_{\textrm{side}} = 8$. We smooth both foreground templates and CMB maps by the Gaussian 
beam of FWHM = $20^{\circ}$. 

\subsubsection{Spectral Index Map}
For spectral index we use {\em WMAP} `best-fit' MCMC 
spectral index map. This map is provided by the {\em WMAP} science team at 
$N_{\textrm{side}} = 64$. We downgrade this map at $N_{\textrm{side}} = 8$ and subsequently 
smooth by a Gaussian beam of FWHM = $20^\circ$. This smoothed spectral index map 
defines our {\it primary} spectral index map, $\beta_s(p)$, for this work with 
which we compare the reconstructed spectral index map obtained by our method.

\subsubsection{Detector Noise Maps and Covariance Matrix}
Generating noise maps and thus modelling the noise properties in the low 
resolution maps are somewhat involved than simple smoothing operations that 
were needed to be performed on foreground templates and CMB maps at 
$N_{\textrm{side}} = 8$ in Section~\ref{ICompSpIndex}. We take into account 
{\em WMAP}'s scan modulated detector noise pattern by using $N_{\textrm{obs}}(p)$ 
information associated with each nine-year Differencing Assembly (DA) map 
available at LAMBDA website at  $N_{\textrm{side}} = 512$. 
We first compute noise variance at each pixel following 
$\sigma^2(p) = \sigma^2_0/N_{\textrm{obs}}(p)$ at $N_{\textrm{side}} = 512$ for all 
DA maps between K to V band. Since for each of Q and V bands 
two DA maps are available we simply average the variances of the corresponding 
DA's at each pixel, to obtain noise variance map corresponding to average DA map 
within the frequency bands.
For {\em Planck} LFI and HFI maps we use simple uniform white noise model. 
Pixel white noise levels, $\sigma(p)$, for {\em Planck} maps are obtained from 
{\em Planck} blue book specifications, $\sigma_{bb}$, which represents 
noise in the beam area of the corresponding map in unit of $\mu K/K$. Pixel 
noise levels for Planck maps at $N_{\textrm{side}}$ are then given by, $\sigma(p) =\sigma_{bb}
T_{CMB}B_{fwhm}/\Delta \theta$, where $B_{fwhm}$ represents 
FWHM of {\em Planck} beams in arcmin, $\Delta \theta=6.9^\prime$, represents pixel 
size of $N_{\textrm{side}} = 512$ map and $T_{CMB}=2.73K$.  

In principle, one can use all the  noise variance maps at $N_{\textrm{side}} = 512$ 
corresponding to all frequency bands of Table~\ref{ListMaps} to obtain 
representations of noise maps at $N_{\textrm{nside}} = 512$ and then downgrade them 
to $N_{\textrm{side}} = 8$ for use in our simulation. However, one needs to simulate 
many such maps at $N_{\textrm{side}} = 512$, since we need to obtain pixel pixel noise 
covariance at $N_{\textrm{side}} = 8$  by downgrading the high resolution noise maps. 
Since generating noise maps at $N_{\textrm{side}} = 512$ and 
subsequently downgrading them many times are computationally 
very expensive we device a completely equivalent approach which is computationally 
significantly fast compared to the former method.  Instead of downgrading 
$N_{\textrm{side}} = 512$ noise maps we obtain noise variance maps at $N_{\textrm{side}} = 8$ 
by simply averaging over noise variance of all $N_{\textrm{side}} = 512$ pixels 
lying inside each $N_{\textrm{side}} = 8$ pixel. Such a method is always feasible 
for our method and does not introduce any pixel-pixel correlation at 
$N_{\textrm{side}} = 8$ since noise in the original $N_{\textrm{side}} = 512$ maps are 
uncorrelated from pixel to pixel. Using these noise variance maps we simulate 
noise maps at $N_{\textrm{side}} = 8$. These noise maps are then smoothed by Gaussian 
beam of FWHM = $20^\circ$, to generate realistic noise maps at $N_{\textrm{side}} = 8$
for each frequency band. We obtain $10000$ Monte-Carlo simulations 
of smoothed noise maps at $N_{\textrm{side}} = 8$. Since the smoothing operations introduces 
pixel to pixel noise covariance we obtain full pixel to pixel noise covariance matrix 
over surviving pixels (after application of {\it pm10} mask) at $N_{\textrm{side}} = 8$ 
corresponding to each frequency band. We have shown noise covariance matrix for V band 
masked map, in Figure~\ref{CovMat}. Non-diagonal nature of this matrix indicates non-trivial 
noise properties in the maps. 

Using the $N_{\textrm{side}} = 8$ smoothed CMB Stokes Q polarized map, $Q_c(p)$, synchrotron 
and thermal dust templates ($Q_{0s}$ and $Q_{0d}$ respectively), 
spectral index map, $\beta_s(p)$ and detector noise map, $Q^n_i(p)$ ($i=1,2,...,n_b$) as 
obtained above we generate simulated frequency maps for {\em WMAP} and {\em Planck} 
frequency bands, using Eqn~\ref{WE1} and  knowing the values of $a_i$. We use 
$\beta_{d} = 2.0$, $\nu_{0s} = 23$ GHz and $\nu_{0d} = 94$ GHz  in this 
work.  

\begin{figure}[t]
\includegraphics[scale=1]{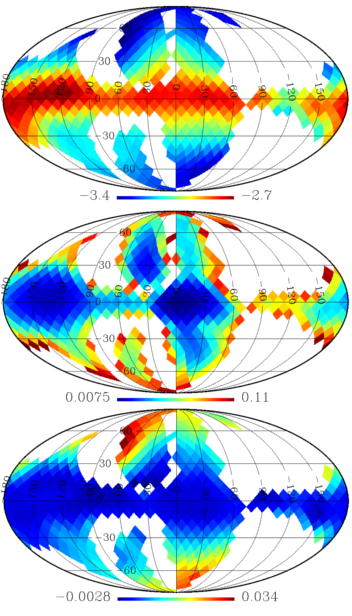}
\caption{{\it Top panel}: Mean synchrotron reconstructed spectral index map obtained 
from $1000$ simulations of this work. {\it Middle panel}: The standard deviation of   
spectral index map. {\it Bottom Panel}: Difference of top panel and input spectral 
index map.}
\label{BetaMaps}
\end{figure}

\subsection{Analysis}
Since spectral indices of neither synchrotron, nor thermal dust components 
are known {\it a priori}, we perform reconstruction of all components defined 
by Eqn~\ref{templates} for a suitable range of both of these indices. We vary 
$\beta_{0s}$ between $-2.65$ to $-3.05$ with a step of $-0.01$. For each 
of these values  $\beta_d$ is varied between $1.80$ and $2.20$ with a 
step size $0.01$. For each pair ($\beta_{0s}, \beta_d$), such generated, we compute shape 
vectors for all foreground template components. Combining these with the shape vector 
of CMB we estimate a set of weights using Eqn~\ref{weights2} to reconstruct 
all template components including CMB. For each pair ($\beta_{0s}, \beta_d$), we  
form reconstructed frequency maps following Eqn~\ref{FreqMapRec} using all 
the reconstructed template components. 
We then form the difference map at each frequency by subtracting  reconstructed 
frequency map from the corresponding input frequency map. 
The noise covariance of the $i^{th}$ difference map over the surviving pixels 
for a given pair of indices is given by the matrix, $N^{i,D}_{pp\prime}$, 
defined by Eqn~\ref{CovDiffMap}. Once the difference map is formed we obtain  
$\chi^2(\beta_{0s},\beta_d)$ for all indices following Eqn~\ref{chi2}. Our 
best fit values for indices are the ones which minimizes $\chi^2$ over all 
indices. Using these values we obtain spectral index variation map following Eqn~\ref{SpIndex}.
We validate our method by performing $1000$ Monte Carlo simulations of  
entire procedure described above. We note that since CMB follows blackbody 
distribution its shape vector remain identical for all foreground indices 
and for all simulations. Since the weights as given by Eqn~\ref{weights2} 
depend only upon foreground and CMB shape vectors (without any dependence
on random detector noise which vary from simulation to simulation) for 
reconstruction purpose of all components, we calculate these weights only once 
for the entire range of spectral indices and  store them on the disk. We then 
use these weights to reconstruct all components for all Monte-Carlo simulations 
for all indices. Moreover, we also precompute Moore-Penrose generalized 
inverses of detector noise covariance matrices of all difference maps for all 
pairs of spectral indices $(\beta_{0s},\beta_d)$ and store them 
on the disk. In effect a Singular Value 
Decomposition algorithm with cutoff $1.0\times 10^{-7}$ was used to implement 
Moore-Penrose inverses.

\section{Results}
\label{Results}

\subsection{Spectral Index Map}
\label{SpIndex:Res}
\begin{figure}[htp]
\centering
\includegraphics[trim=10 10 0 90, clip,scale=0.7]{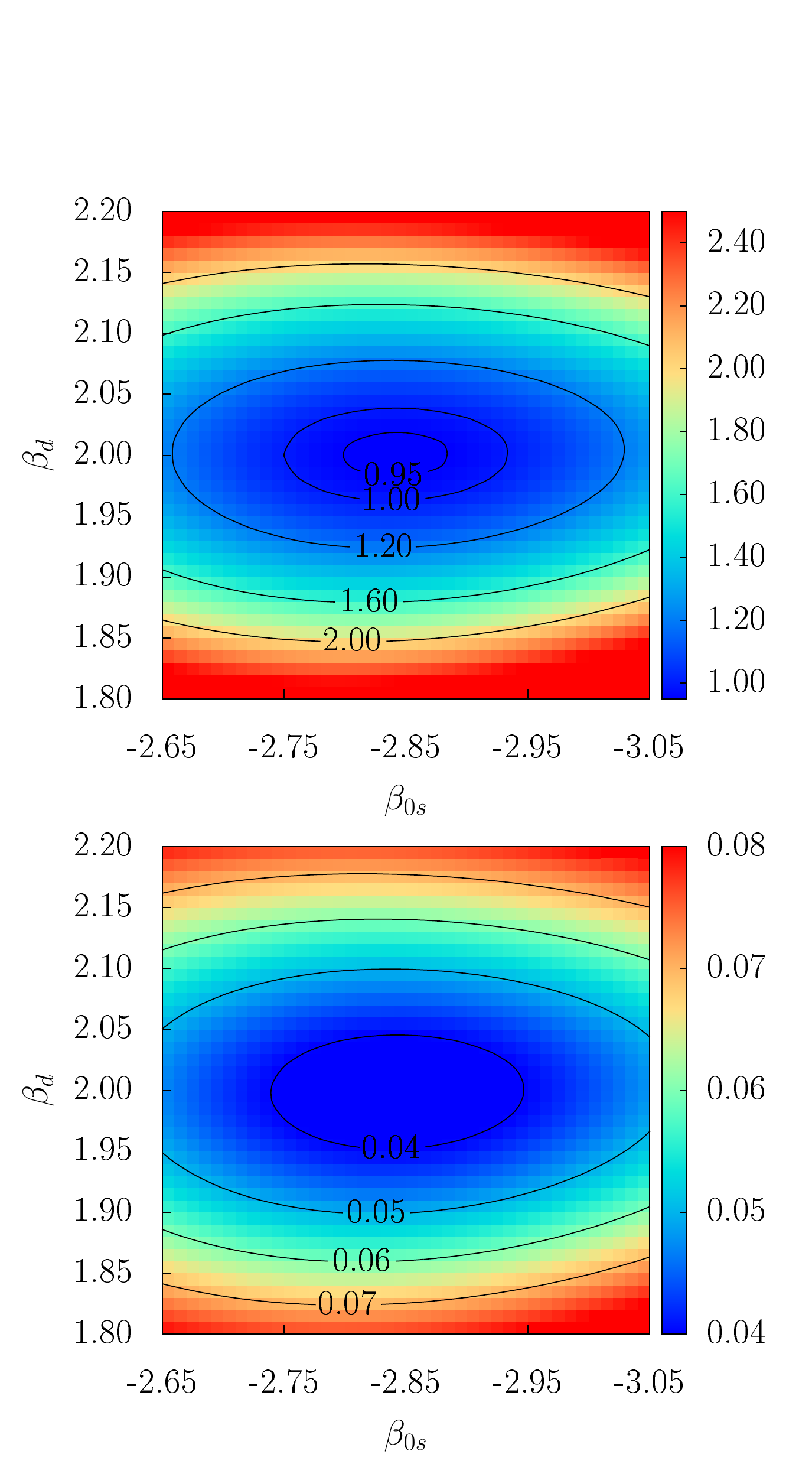}
\caption{{\it Top panel:} Variation of mean $\chi^2(\beta_{0s},\beta_d)$  
with respect to spectral indices obtained from $1000$ Monte Carlo simulations. 
{\it Bottom panel:} Standard deviation map of $\chi^2(\beta_{0s},\beta_d)$ 
for any one of the $1000$ simulations.}
\label{Chi2Plot}
\end{figure}

For each simulation described in Section~\ref{Methodology}, we construct 
spectral index variation map, $\Delta \hat \beta_s(p)$, over the reference index, 
$\beta_{0s}$, following Eqn~\ref{SpIndex}. 
The results for reconstruction of spectral index map are shown in Figure~\ref{BetaMaps}. The colored 
area of Mollweide projection of all the three panels shows the region on the 
sky that survives after applying {\it pm10} mask. The top panel shows mean reconstructed 
synchrotron spectral index map, $\left<\hat\beta_s(p)\right>$, defined as, 
$\left<\hat\beta_s(p)\right>= \left <\beta_{0s,\textrm{min}}\right>+
\left<\Delta \hat \beta_s(p)\right>$, where $\beta_{0s,\textrm{min}}$ denotes 
value of $\beta_{0s}$ (e.g., see Eqn~\ref{WE3}) corresponding to $\chi^2_{min}$ and the 
average is performed over all $1000$ simulations. The middle 
panel is simple standard deviation map for any one of the reconstructed spectral 
index maps. The lower most panel shows the difference between average reconstructed 
index map and the input spectral index map, $\left< 
\hat \beta_s(p)\right>-\beta_s(p)$. Clearly, the difference map shows our 
method can accurately reconstruct spectral index map inside the {\it pm10} mask. 
The reconstruction difference towards higher (lower) galactic coordinates becomes 
slightly larger due to weak polarized signal in these regions.

\begin{figure}[t]
\includegraphics[scale=0.7]{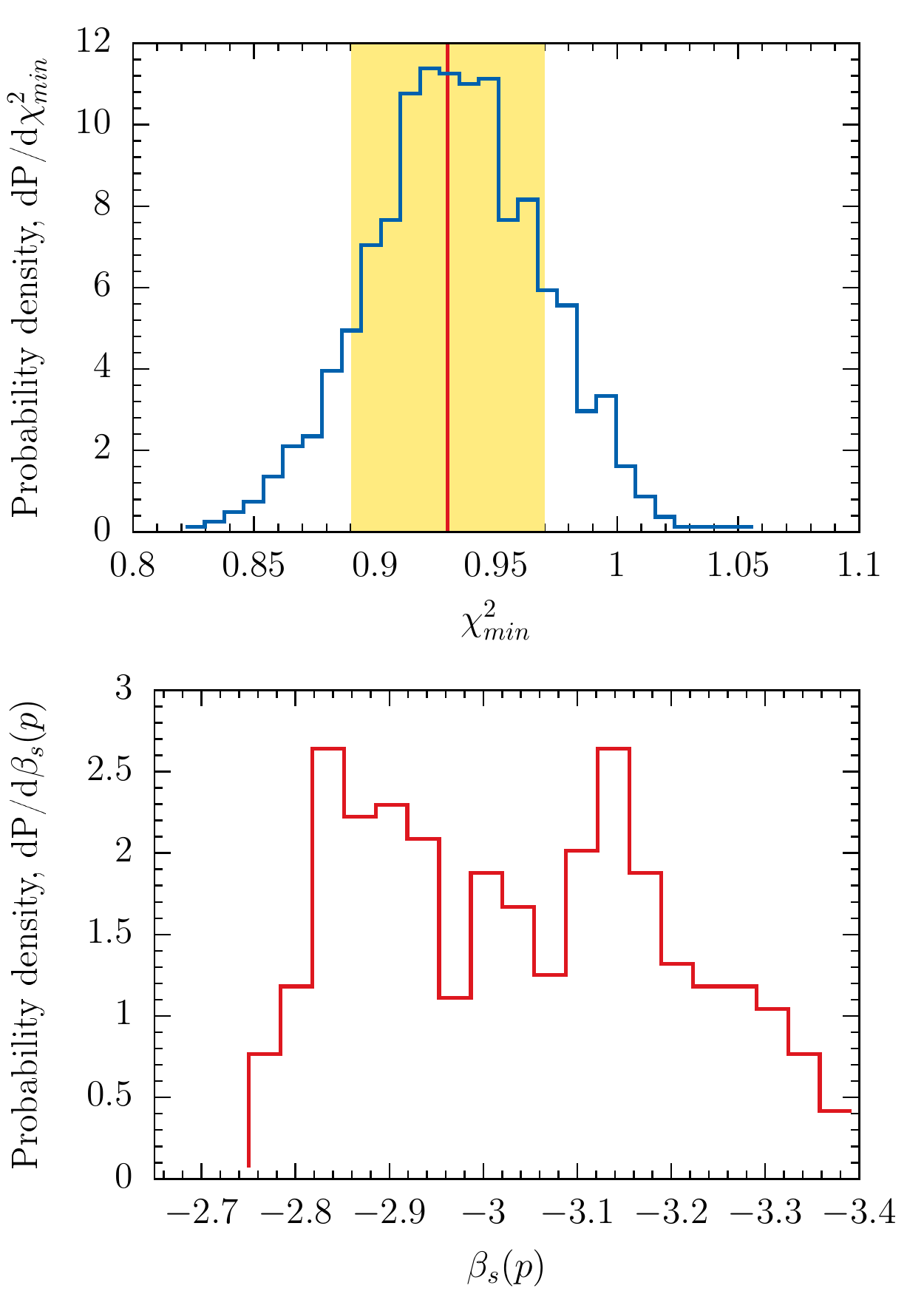}
\caption{{\it Top panel:} Probability density of minimum $\chi^2$ values 
obtained from $1000$ simulations. Red vertical line shows the location of sample 
mean corresponding to $\chi^2_{min}=0.93$. Light golden band around the mean 
value shows regions within 1$\sigma$ deviation. {\it Bottom panel:} Probability
density function of pixel indices obtained from smoothed input spectral index map after
application of {\it pm10} mask. }
\label{chi2pdf}
\end{figure}

\begin{figure*}[t]
\includegraphics[trim=70 20 0 10,clip,scale=0.7]{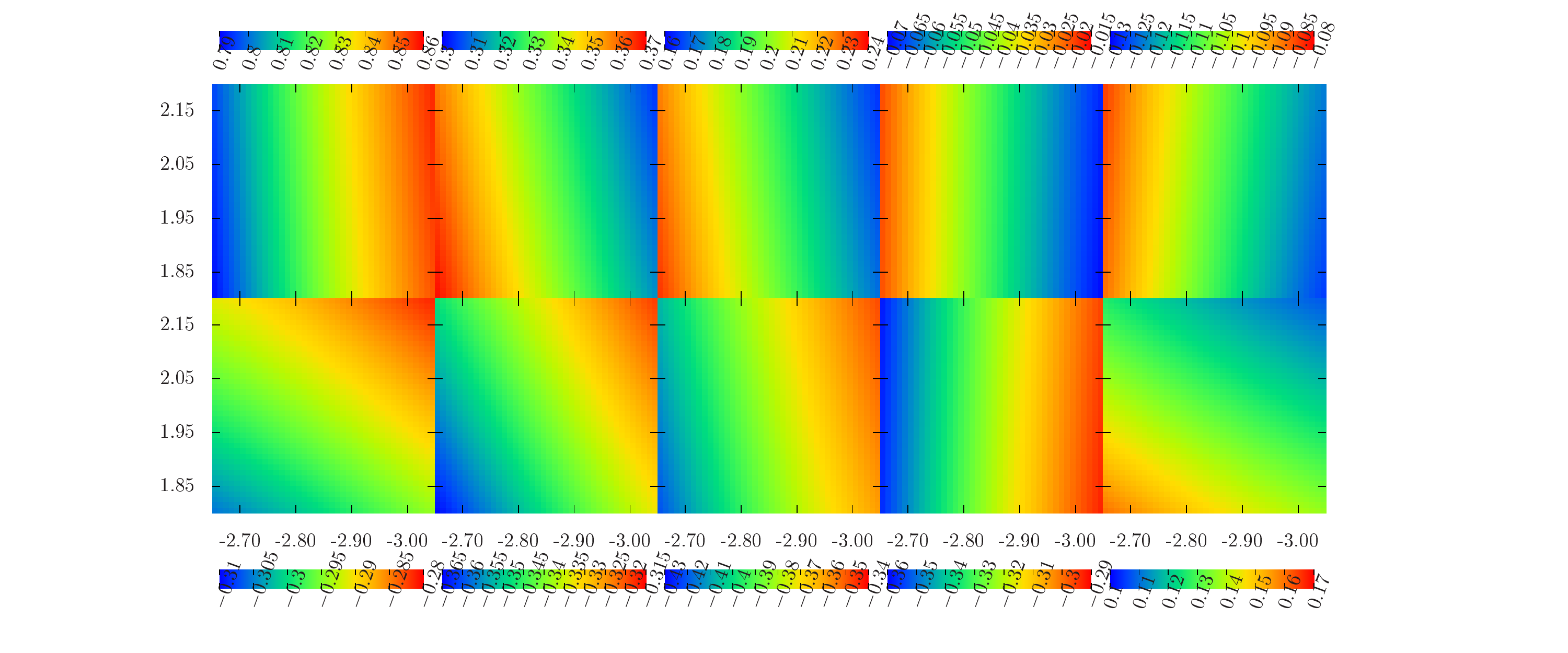}
\includegraphics[trim=70 20 0 10,clip,scale=0.7]{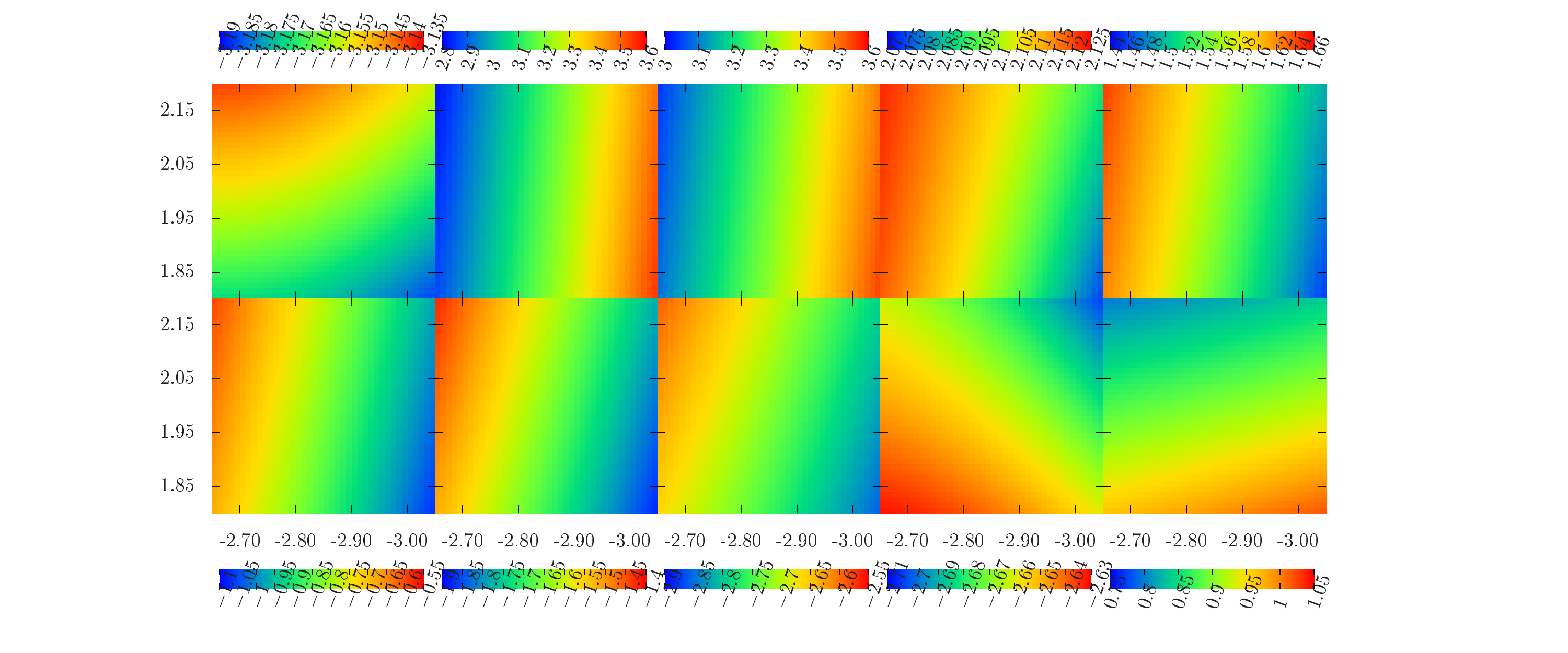}
\caption{{\it Top panel:} Weights for reconstruction of $Q_{0s}(p)$. {\it Bottom panel:} Weights 
for reconstruction of $Q_{0s}(p)\Delta \beta_s(p)$. For any given panel, plots in consecutive rows 
from left to right indicate respectively weights for K, A, Ka, Q, B, C, V, D, E and F frequency bands.}
\label{WeightMap}
\end{figure*}

\begin{figure*}[t]
\includegraphics[scale=1.4]{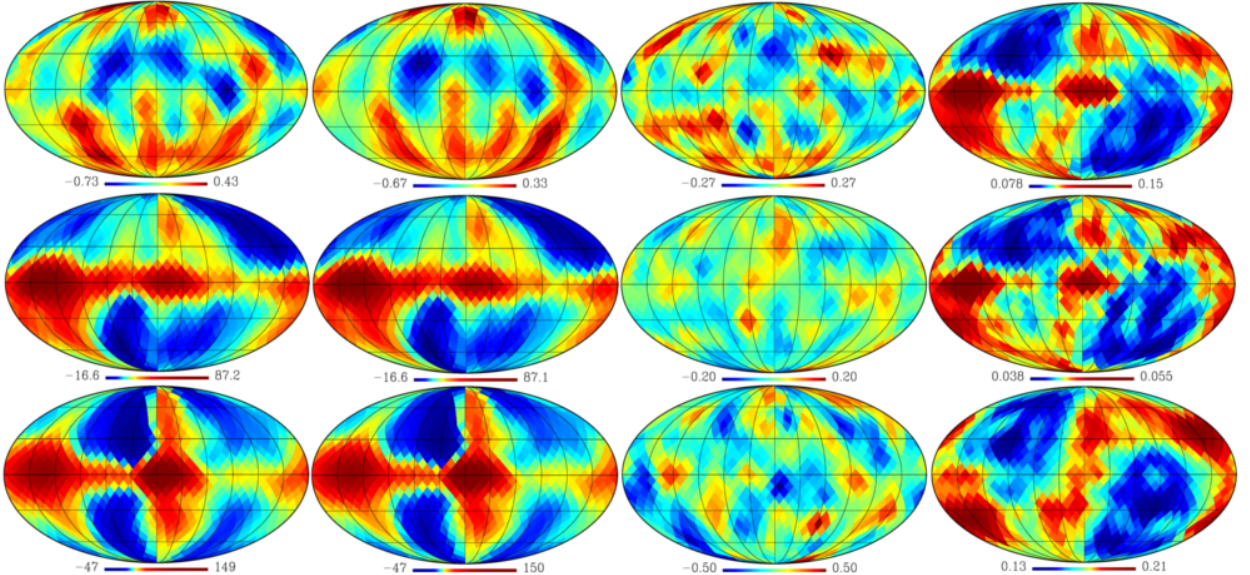}
\caption{Results for ML template reconstruction for CMB (top panel), thermal dust (middle panel)
and synchrotron (bottom panel). Units for all colormaps are in $\mu K$ temperature unit (thermodynamic
for CMB, antenna for foreground components).}
\label{compmaps}
\end{figure*}

For any given Monte-Carlo simulation  $\chi^2(\beta_{0s},\beta_d)$ values shows 
a clear minimum, $\chi^2_{min}$, at some values of $(\beta_{0s},\beta_d) = (\beta_{0s,\textrm{min}},\beta_{d,\textrm{min}})$. Although
a topographical plot of $\chi^2(\beta_{0s},\beta_d)$ depends upon the particular 
random simulation under consideration, the features of such plots that arise on the average due to 
foregrounds,  CMB  and detector noise, can be studied by making a plot of $\left<\chi^2(\beta_{0s},\beta_d)\right>$
(where the average is performed over all $1000$ simulations for each pair $(\beta_{0s},\beta_d)$) with respect to both indices.      
We have shown this at the top panel of Figure~\ref{Chi2Plot}. The elongated contours 
of this plot shows the relatively higher sensitivity of the data with the variation of synchrotron spectral
index compared to the same variation of dust spectral index. In the bottom panel of this 
figure we show standard deviation map of $\chi^2$ values for all indices. Lower standard 
deviation values around $\left<\chi^2_{min}\right>$ indicates minimum is recovered 
satisfactorily by our method. Using results from Monte Carlo simulations we estimate 
average value of $(\beta_{0s,\textrm{min}},\beta_{d,\textrm{min}})$ corresponding to $\chi^2_{min}$ as 
$(-2.84 \pm 0.01, 2.00 \pm 0.004)$. Small error on the spectral index values
corresponding to $\chi^2_{min}$ shows that our method can reliably recover global
spectral indices.

Using the Monte Carlo simulations we find that values of $\chi^2_{min}$ is distributed 
over a narrow range of $0.82$ to $1.05$. A plot of histogram of $\chi^2_{min}$  is 
shown in top panel of Figure~\ref{chi2pdf}. We estimate, $\left<\chi^2_{min}\right> = 0.93$, 
averaging over  $1000$ Monte-Carlo simulations, with sample standard deviation  $0.04$.
The probability density function of this figure, follows a very symmetric distribution for $\chi^2_{min}$.
This can be  expected due to very large number of degrees of freedom ($\sim 4260$) available in the analysis.
In the bottom panel of this figure we have shown probability distribution of $20^\circ$
smoothed $N_{\textrm{side}} = 8$ input spectral index map, from the region that survives 
after application of $\it pm10$ mask. A bimodal structure of the distribution is readily 
identifiable corresponding to spectral indices $\sim -2.85$ and $-3.16$ respectively. 
Comparing these values with the mean reconstructed spectral index $-2.84 \pm 0.01$, it is 
interesting to note that our method picks out the  peak corresponding to a flatter index,
from available, a pair of  doubly degenerate ones.  
 
\subsection{Weight}
One very important component of this work are the weights. We show weights for reconstruction of 
$Q_{0s}(p)$ and $Q_{0s}(p)\Delta \beta_s(p)$ templates for the entire range of spectral 
indices, respectively in the top and bottom panel of Figure~\ref{WeightMap}.  At the low 
frequency side, where synchrotron emission is dominant weights tend to be 
positive for all indices. We see this for K, A and Ka bands. For all other bands 
except F band, weights are negative for all indices. For F band weights become positive 
for all indices to cancel out signals from CMB, thermal dust,  $Q_{0s}(p)\Delta \beta_s(p)$
and at the same time to preserve $Q_{0s}$. For reconstruction of $Q_{0s}(p)\Delta \beta_s(p)$ template
weights must completely project out  its shape vector from the data, at the same time, the weight vector  need to 
be  orthogonal to shape vectors of synchrotron and other components. Because of close analytical
similarity in the shape vectors of $Q_{0s}(p)$ and $Q_{0s}(p)\Delta \beta_s(p)$ the orthogonality 
condition leads to somewhat larger magnitude of weights for $Q_{0s}(p)\Delta \beta_s(p)$ than  $Q_{0s}(p)$
template.  Relatively larger magnitude of weights can easily be seen from the second panel
than the first one. For the second panel and for K band, weights take largest negative value, since, from
the expression of shape 
vector for $Q_{0s}(p)\Delta \beta_s(p)$ component, ($a(\nu)(\nu/\nu_{0s})^{-\beta_{0s}}\ln(\nu/\nu_{0s})$),
shape vector has a zero component  at K band. For low frequency A, Ka, Q, B bands all weights are positive 
indicating its larger shape vector at lower frequency. For V, C, D and E bands weights become 
negative, since shape vector decreases in this frequency range. Finally for F band weights again become 
positive to project out $Q_{0s}(p)\Delta \beta_s(p)$ component completely and to cancel out rest of the components.

\subsection{CMB and Foreground}
\label{CMBForeground}
As associated results of our method, we reconstruct  ML templates 
for all components using the best fit spectral index values 
$\beta_{0s,\textrm{min}},\beta_{d,\textrm{min}}$ corresponding 
to $\chi^2_{min}$ obtained in Section~\ref{SpIndex:Res}. We show the results in Figure~\ref{compmaps}. 
Unlike the case for reconstructed synchrotron spectral map, we 
do not apply any mask on the recovered templates, since CMB and 
foreground signals are expected to be stronger than the spectral
index variations. Top, middle and bottom panels of this figure show results for 
CMB, thermal dust and synchrotron template reconstruction 
respectively, obtained from a randomly chosen Monte Carlo 
simulation with seed = $100$. All the color maps of this figure 
are in $\mu K$ unit (thermodynamic for CMB, antenna for foreground 
templates). Following usual row, column convention of a matrix, we 
represent any image of this figure by the pair $(i,j)$. $(1,1)$ image 
represents reconstructed CMB map against the input CMB map for the 
same simulation, shown as $(1,2)$ image. As one can easily visually 
identify the large scale features of these two maps are very nearly 
identical. $(1,3)$ image shows the difference between 
reconstructed and input CMB map and hence is dominated by residual 
detector noise. The noisy nature of the difference map is also 
indicated by the symmetric color table plotted below $(1,3)$ image. 
The $(1,4)$ image represents the standard deviation map estimated 
from set of all $(1,3)$ maps resulting from $1000$ Monte Carlo 
simulations of component separation algorithm. As expected, the 
standard deviation map clearly indicates a scan modulated noise 
pattern as applicable for {\em WMAP} observations. There exist 
some foreground induced errors on the galactic plane. Such errors 
can be reduced by extending our method to a second order perturbative 
analysis.   

The $(2,1)$ image represents reconstructed thermal dust template 
at $217$ GHz for  seed = $100$. The reconstructed template matches very 
well with the input thermal dust template at the same
frequency, viz., $(2,2)$ image. The difference between reconstructed 
and input template ($(2,3)$ image) is dominated by noise. The standard 
deviation map for reconstructed thermal dust template is shown 
as $(2,4)$ image. Clearly, {\em WMAP}'s scan modulated noise pattern is 
visible in this figure. There also exist some residual foregrounds error  
in this map, which again can be eliminated 
by performing a higher order perturbative analysis.

The $(3,1)$ image shows  reconstructed synchrotron template at $23$ GHz 
for the same random seed. The input synchrotron template at $23$ GHz is 
shown as $(3,2)$ image. Both these maps match with each each other very 
well. The difference of the first and second maps is shown as (3,3) image. 
The standard deviation map, image $(3,4)$, shows the scan modulated 
detector noise pattern,  without any visible signature of residual 
foreground error.

\section{Discussion \& Conclusion}
\label{Conclusion}

Reconstruction of CMB and foreground components jointly from microwave observations 
is a primary task for cosmologists. This can be achieved relatively easily if i)
observed sky signal consists of only those components that have rigid frequency scaling all 
over the sky ii) detector noise is negligible and iii) total number of these components 
is less than or equal to total number of observed frequency bands. However, the task is 
complicated when spectral index of any one of the components varies with sky 
positions. The stronger the variation effectively one gets more components with 
rigid frequency scaling, which, if  not taken into account in the analysis, may cause 
the results to be biased, or at the least difficult to interpret. The scenario
becomes even worse in the presence of non-negligible detector noise. {\it We have shown
that in the presence of foregrounds following  power law model with spatially varying 
index and small amount of detector noise one can recover the spectral index variation
map reliably from the {\it WMAP} and {\it Planck} Stokes Q observations without any need to model
foregrounds and or CMB in terms of any model templates.} The only assumption 
we have made in this work is a power law model for the foregrounds. However, this is 
not a {\it necessary} condition for our method to work, since one can directly 
fit for the shape vectors for all components by expanding the parameter space.  Then with the help 
of perturbative expansion of the spectral index variation one can reconstruct the 
index variation map. Since the spectral index variation for synchrotron component 
is small compared to other emissions (specifically, compared to synchrotron emission 
amplitude itself), and can be masked by the presence of comparable level of  detector noise 
contamination. we chose to work with maps containing low level of detector noise, by smoothing 
the frequency maps by a $20^{\circ}$ beam window function.
We use full pixel-pixel detector noise covariance matrix to accurately model the smoothed 
detector noise pattern, and thus for better reconstruction of spectral index map. The 
condition for large beam window can be relaxed easily for experiments, or for sky regions 
with negligible detector noise.  A very interesting future project will be to apply the 
method on the {\it Planck} polarized observations combined with {\it WMAP}, and with  
polarization sensitive, low noise, future CMB observations like CMBPol. {\it For very low
noise experiments in future, it would be very useful to extend our method up-to second 
order perturbative analysis for even more accurate understanding of cosmological and 
astrophysical information contained in data.}

We summarize main results and conclusions of this work as follows: 
\begin{enumerate} 
\itemsep0em
\item {We  present and implement a first order perturbative method 
to estimate variation of synchrotron spectral index over the microwave 
sky using simulated Stokes Q observations of {\it WMAP} K, Ka, Q, V, 
{\it Planck} LFI, and three lowest frequency {\it Planck} HFI frequency
bands. We validate the method by performing $1000$ Monte Carlo 
simulations over regions of the sky dominated by synchrotron Stokes Q 
signal. Our results suggest that, following the first order analysis 
synchrotron  spectral index variation map can be reconstructed 
reliably by our method with a narrow error margin.
}
\item {As accompanying results of this work, we  reconstruct CMB, 
thermal dust and synchrotron template components. An error 
estimation on the reconstructed templates suggest that the template 
components can be reconstructed reliably by our method.} 
\item {Our reconstructed templates and best fit spectral indices 
are ML estimates.}
\item {The perturbative method of this work possesses an interesting 
property that it can be tuned to the data with an accuracy, as allowed 
by the data. This could be particularly very useful for CMB and foreground
component separation problem if the data is insufficient to constrain an
otherwise completely accurate model, due to underlying degeneracies 
in the foreground components. }
\end{enumerate}

The work presented in  this paper will allow an improved estimation 
of cosmological signal due to its detailed foreground modelling in terms 
of variation of spectral indices. It will be useful to extract astrophysical 
information about galactic synchrotron emission, in particular, information 
about the galactic  cosmic ray electrons and  the magnetic field. Finally,
the method can be used to extract jointly all foregrounds components along 
with CMB signal from  CMB observations.
  
RS acknowledges research grant SR/FTP/PS-058/2012 by SERB, DST.  Some of 
the results of this paper were obtained using publicly available 
HEALPix~(\cite{Gorski2005}) package available from web page {http://healpix.sourceforge.net}. 
We acknowledge the use of the Legacy Archive for Microwave Background Data 
Analysis (LAMBDA), part of the High Energy Astrophysics Science Archive 
Center (HEASARC). HEASARC/LAMBDA is a service of the Astrophysics Science 
Division at the NASA Goddard Space Flight Center.

\bibliographystyle{apj}

\end{document}